\documentclass[%
 reprint,
 amsmath,amssymb,
 aps,
]{revtex4-1}

\usepackage{graphicx}
\usepackage{dcolumn}
\usepackage{bm}
\usepackage{color}
\usepackage{mathtools}
\usepackage{rotating}
\usepackage{multirow}
\usepackage[table,xcdraw]{xcolor}

\begin{document}

\title{Apparent superballistic dynamics in one-dimensional random walks with biased detachment}

\author{Chapin S.\ Korosec, David A.\ Sivak, Nancy R.\ Forde}
\email{ckorosec@sfu.ca, dsivak@sfu.ca, nforde@sfu.ca}
 \affiliation{Department of Physics, Simon Fraser University, 8888 University Drive, Burnaby, British Columbia, V5A 1S6, Canada}
\date{\today}

\begin{abstract}
	The mean-squared displacement (MSD) is an averaged quantity widely used to assess anomalous diffusion. 
    In many cases, such as molecular motors with finite processivity, dynamics of the system of interest produce trajectories of varying duration.
    Here we explore the effects of finite processivity on different measures of the MSD.
	We do so by investigating a deceptively simple dynamical system: a one-dimensional random walk (with equidistant jump lengths, symmetric move probabilities, and constant step duration) with an origin-directed detachment bias. By tuning the time dependence of the detachment bias, we find through analytical calculations and trajectory simulations that the system can exhibit a broad range of anomalous diffusion, extending beyond conventional diffusion to superdiffusion and even superballistic motion.
	We analytically determine that protocols with a time-increasing detachment
	lead to an ensemble-averaged velocity increasing in time, thereby providing the effective acceleration that is required to push the system above the ballistic threshold. MSD analysis of burnt-bridges ratchets similarly reveals superballistic behavior. Because superdiffusive MSDs are often used to infer biased, motor-like dynamics, these findings provide a cautionary tale for dynamical interpretation. 
\end{abstract}

\maketitle

\section{Introduction}

The mean-squared displacement (MSD) is often used to assess anomalous diffusion in molecular systems. A system with an MSD that grows linearly in time is conventionally diffusive, lacking anomalous effects~\cite{Einstein1956}: 
\begin{equation}\label{linDiff}
{\mathrm{MSD} = \left\langle x^{2}(t) \right\rangle  = 2Dt .}
\end{equation} 
For processes that obey \eqref{linDiff}, such as a discrete-time random walk, the displacement distribution after $N$ steps limits to a Gaussian distribution. That is, the $N$ individual steps of a discrete-time random walk are independent and identically distributed, with their sum governed by the central limit theorem~\cite{Metzler2000}. 

Anomalous diffusion refers to systems that do not have a linear time dependence of the MSD. Generally, anomalous diffusion is thought to emerge in stochastic systems whose displacement distributions are not Gaussian, and is therefore intimately connected with the breakdown of the central limit theorem~\cite{Metzler2000}. The concept of anomalous diffusion was first introduced in 1926 by L.F.\ Richardson~\cite{Richardson1925} through a thought experiment involving two independent air particles separated by a sufficiently large distance so as to be caught by two independent gusts of wind moving in opposite directions. Richardson hypothesized that such a system does not obey Fick's second law, and that the MSD of the air particles scales nonlinearly with time. For an anomalously diffusive system,
\begin{equation}\label{anoDiff}
\mathrm{MSD} \propto
D_{\rm g} t^{\alpha} \ ,
\end{equation} 
where $D_{\rm g}$ is the generalized diffusion coefficient, and $\alpha$ is the anomalous diffusion exponent that distinguishes the type of diffusion~\cite{Havlin2002, Metzler2000, Metzler2014}. Subdiffusion, conventional diffusion, and superdiffusion correspond to $0 \leq 
\alpha < 1$, $\alpha = 1$, and $1 < \alpha < 2$, respectively. The ballistic threshold is at $\alpha = 2$, describing a system whose trajectories proceed at constant velocity.

In order for a microscopic system to behave ballistically over long times an external stimulus is typically required. For example, ballistic motion is achieved for random walkers subject to certain forms of external noise~\cite{Bao2003}.
Ballistic motion can also be achieved without an external stimulus in systems whose cooperative behavior limits the degrees of freedom of individual particles~\cite{Feinerman2018, Rank2013}. 
To exceed $\alpha = 2$, reaching the superballistic regime, it is generally thought an acceleration is required~\cite{Redner1992}. 
Examples include animal movement generated by muscle contraction~\cite{Tilles2017} and particles optically trapped in air~\cite{Li2010,Duplat2013}, subject to increasing temperatures~\cite{Cherstvy2016}, or subject to expanding media~\cite{LeVot2017}. 

In this work we consider a deceptively simple one-dimensional discrete random walk with equal-sized and equal-duration steps.
We impose no external forces, particle thrust, noise typically thought to promote superdiffusive motion, or cooperative behavior. We instead impose a tunable detachment probability $d$ that produces finite processivity. 
For every step toward its initial position, the random walker permanently detaches with probability $d$.
We explore two detachment protocols: constant detachment, and detachment exponentially increasing at rate $k_{\rm d}$.

By varying $d$ or $k_{\rm d}$, we find that the anomalous diffusion exponent $\alpha$ can be tuned over a wide range of values.
Varying $d$ controls apparent time-dependent dynamics ranging from diffusive to ballistic, while variations in $k_{\rm d}$ tune $\alpha$ from the diffusive into a transient but long-lasting superballistic regime.
We determine the mechanism for this apparently superballistic behavior by analytically deriving the ensemble velocity and find it is dependent on the detachment protocol used, thereby producing the ensemble acceleration required to breach the ballistic threshold. 
We reproduce several of these observations in a trajectory-resolved model,
though find that distinct approaches to calculating the MSD provide different values of $\alpha$ and hence distinct inferences about the system's dynamics. 
We additionally show that selecting a subset of especially processive trajectories---as is common in single-particle studies~\cite{Salman2002, Howse2007,Manley2008}---can introduce a large superdiffusive bias into the MSD.
As an example, we demonstrate that this acceleration-by-detachment can manifest in a more realistic system by calculating MSD for burnt-bridges ratchets, which we find to exhibit apparently superballistic behavior. Our results encourage increased transparency regarding selection methods for single-particle analysis.

\section{Modeling random walks with detachment}
\subsection{Ensemble-level model} 

First we study a discrete-time diffusion scheme where we track the probability flow of an ensemble of independent one-dimensional random walkers.
At time intervals $\Delta t = 1$, each walker moves a distance $\Delta x = 1$, with equal probability of going left or right.
A step towards the initial position at the origin incurs a probability of detachment, either with constant or with exponential probability. 
For constant detachment, the probability $d$ of detaching is independent of time.
Exponential detachment probability is given by  
\begin{equation}\label{det}
d(t) = 1 - e^{-k_{\rm d}t} \ ,
\end{equation}
increasing over time as determined by the rate $k_{\rm d}$. 
(Thus the probability of remaining attached at each origin-directed step is $r(t) = 1 - d(t)$.) Detachment events of individual walkers are independent.
Table~\ref{table1} shows such a system's initial evolution.
For the results shown in this work, the probability distribution was evolved for $n = 1000$ timesteps.

\begingroup
\renewcommand{\arraystretch}{1.3}
\begin{table}[h]
\caption{Probability distributions for the first 4 time steps of the ensemble-level model.}
\begin{tabular}{ccccccccccc}
\multicolumn{1}{l}{}  &                        & \multicolumn{9}{c}{Displacement} \\
\multicolumn{1}{l}{}  & \multicolumn{1}{c|}{}  & -4 & -3 & -2 & -1 & 0 & 1 & 2 & 3 & 4 \\ 
\cline{2-11} 
\multirow{5}{*}{\begin{sideways}{Time}\end{sideways}} & \multicolumn{1}{c|}{0} & & & & & 1 & & & & \\
& \multicolumn{1}{c|}{1} & & & & $\frac{1}{2}$ & 0 & $\frac{1}{2}$ & & & \\
& \multicolumn{1}{c|}{2} & & & $\frac{1}{4}$ & 0 & $\frac{1}{2}r$ & 0 & $\frac{1}{4}$ & & \\
& \multicolumn{1}{c|}{3} & & $\frac{1}{8}$ & 0 & $\frac{3}{8}r$ & 0 & $\frac{3}{8}r$ & 0 & $\frac{1}{8}$ & \\
& \multicolumn{1}{c|}{4} & $\frac{1}{16}$ & 0 & $\frac{1}{4}r$ & 0 & $\frac{3}{8}r^{2}$ & 0 & $\frac{1}{4}r$ & 0 & $\frac{1}{16}$
\end{tabular}
\label{table1}
\end{table}
\endgroup

\subsection{Trajectory-resolved model}\label{TrajSpec}

We model a one-dimensional discrete-time random walk by sampling step lengths from a Gaussian distribution with zero mean and unit standard deviation.
Steps toward the origin lead to detachment with probability described by \eqref{det}.
As a control, we compare to a random walk with no detachment, which produces conventional diffusion.

\section{Analysis Methods}
\subsection{Mean-squared displacement and $\alpha(t)$ for ensemble-level model}

We define the MSD for the ensemble-level model as
\begin{equation}\label{rem}
\mathrm{MSD_{\rm EA}} = \left\langle x^{2} \right\rangle _{\rm rem} \equiv \frac{\sum p_{i} x_{i}^{2}}{P_{\rm rem}},
\end{equation}
where $x_{i}$ is the $i^{\rm th}$ position, $p_{i}$ is the probability of having a walker at the $i^{\rm th}$ site, and $P_{\rm rem}$ is the total remaining probability. MSD$_{\rm EA}$ denotes an ensemble-averaged MSD, discussed more in the following sub-section.

The time dependence of the anomalous diffusion exponent $\alpha$ is sometimes calculated through a quantity called either the \emph{dynamic functional}~\cite{Kepten2013} or the \emph{local MSD scaling exponent}~\cite{Ghosh2016}, which is the derivative of the logarithm of the MSD with respect to the logarithm of time~\cite{Karmakar2016}.
\begin{equation}\label{dynamicF}
\alpha_{\psi} \equiv \frac{\mathrm{d(\log MSD)}}{\mathrm{d}(\log \Delta t)} \ . \end{equation}
It has been noted that the numerical derivative of \eqref{dynamicF} can become especially noisy at long times because of compression of linear sampling by logarithmic scaling, thereby introducing uncertainty into estimates of the time dependence of $\alpha$~\cite{Martin2002}.

Alternatively, we introduce
a two-point estimator $\alpha^{n}_{n-b}$ for $\alpha$ using timepoints $n$ and $n-b$. 
In one dimension and for no average drift ($\left\langle x \right\rangle = 0$), 
\begin{equation}\label{alphaMAIN}
\alpha^{n}_{n-b} \equiv \log_{n} \left( \frac{\left\langle x^{2} (n) \right\rangle}{\left\langle x^{2} (n-b) \right\rangle} \right) \left( 1 - \frac{1}{\log_{n-b} n}\right)^{-1} \ .
\end{equation}
(Appendix A provides a general derivation.) To the best of our knowledge, \eqref{alphaMAIN} has not been used before, and we find it useful for characterizing 
transient anomalous diffusion 
with 
the MSD. It is important to note that $b$ must be kept small in order to extract information about the local slope; for this reason we use $b = 2$ in this work when estimating the time dependence of $\alpha$ with \eqref{alphaMAIN}. 

As an example of its utility, in Appendix B we compare $\alpha^{n}_{n-b}$ to $\alpha_{\psi}$ when the MSD undergoes an abrupt change in slope. Figure~\ref{comparison} shows that the numerical derivative from \eqref{dynamicF} produces a noisy approximation to $\alpha_{\psi}$, whereas $\alpha^{n}_{n-b}$~\eqref{alphaMAIN} is smooth throughout. 

Table~\ref{table2} in Appendix C presents $P_{\rm rem}$, $\sum p_{i} x_{i}^{2}$, $\left\langle x^{2} \right\rangle_{\rm rem}$, and $\alpha^{n}_{n-2}$ for the first 4 timesteps.

\subsection{Mean-squared displacement and $\alpha(t)$ for trajectory-resolved model} 

There are many ways to compute the MSD of trajectories; we start by defining the squared displacement as
\begin{equation}\label{sqDisp}
\Delta x_{j}^{2} \left(t, \tau \right)  \equiv \left[ x_{j}(t + \tau) - x_{j}(t)\right]^{2} \ ,
\end{equation}
where $x_{j}$ denotes the position of the $j$th trajectory, $t$ is time, and $\tau$ is the time lag.   
The mean-squared displacement for an ensemble of $N$ independent walkers is
\begin{equation}\label{EAMSD_general}
\left\langle {\Delta x^{2} \left(  t, \tau \right) }\right\rangle \equiv \frac{1}{N} \sum_{j=1}^{N}\Delta x_{j}^{2} \left( t, \tau \right) \ .
\end{equation}

We define the ensemble-averaged (EA) MSD as the MSD at time $\tau$, relative to the initial time zero: 
\begin{equation}\label{EAMSD}
\mathrm{MSD_{\rm EA}} \equiv \left\langle {\Delta x^{2} \left(  0, \tau \right) }\right\rangle  = \frac{1}{N} \sum_{j=1}^{N}\Delta x_{j}^{2} \left( 0, \tau \right) \ .
\end{equation}
This is equivalent to~\eqref{rem}, above. By contrast, the trajectory-averaged (TA) MSD is computed independently for each trajectory,
\begin{equation}\label{TAMSD}
\mathrm{MSD_{\rm TA}} \equiv \overline{  \Delta x_{j}^{2} \left( \tau \right) }  = \frac{\Delta t}{T_{j} - \tau + \Delta t}  \sum_{t = 0}^{T_{j} - \tau}\Delta x_{j}^{2} \left( t, \tau \right) \ ,  
\end{equation}
where $T_{j}$ is the total duration of the $j$th trajectory, and $\Delta t=1$ is the time step.

The \emph{van Hove correlation function} or \emph{displacement distribution} $G(x, \tau)$ collects all the displacements for a particular time lag $\tau$ across all the observed trajectories~\cite{vanHove1954, Akcasu1970,Ghosh2016}. $G(x, \tau)$ describes the probability of a particular displacement for a time lag $\tau$. The variance of $G(x, \tau)$ is the ensemble MSD$_{\rm TA}$ (\emph{i.e.}, the MSD$_{\rm ETA}$) for $\tau$:
\begin{equation}\label{ETAMSD}
\mathrm{MSD_{\rm ETA}} \equiv \left\langle\overline{\Delta x^{2} \left( \tau \right) }\right\rangle   = \frac{1}{N}\sum_{j = 1}^{N} \overline{\Delta x_{j}^{2} \left( \tau \right)} \ .
\end{equation}
The MSD$_{\rm ETA}$ is particularly useful for many independent but short trajectories. 

MSDs calculated from equations \eqref{EAMSD}, \eqref{TAMSD}, and \eqref{ETAMSD} are asymptotically equal for an ergodic process: the time average converges to the ensemble average at a sufficiently long time.  
By contrast, differing MSDs may imply nonergodicity; examples of such systems include tracer particles in mucus~\cite{Cherstvy2019} and a continuous-time random walk with diverging average waiting times~\cite{Bel2005}. 

Here, we estimate $\alpha$ through a linear fit of log MSD as a function of log time. For visual aid we fit over local regions of approximate linearity and report the range of times used in the fit. 
Because this logarithmic scaling produces a non-normal distribution of errors, we use a generalized least squares (GLS) approach to estimate $\alpha_{\rm GLS}$ from the MSD~\cite{Orsini2009}, employing the function \textit{gls} from the \textbf{R} library \textit{nlme}. 

\subsection{Shape of probability distributions}

Deviation of the displacement distribution $G(x, \tau)$ from a Gaussian can be quantified by the standardized fourth moment (the kurtosis), 
\begin{equation}\label{kurt}
\beta_{2}(x) \equiv \frac{\left\langle (x - \left\langle x \right\rangle )^{4} \right\rangle }{(\left\langle (x - \left\langle x \right\rangle )^{2}\right\rangle )^{2}} \ .
\end{equation}
As a Gaussian distribution has $\beta_{2}(x) = 3$, it is convenient to define the excess kurtosis $\gamma_{2} \equiv \beta_{2} - 3$, such that any non-zero $\gamma_{2}$ indicates non-Gaussian behavior~\cite{Rahman1964,kob1995}.

\section{Results \& Discussion}
\subsection{Ensemble-level analysis}

\begin{figure}[h]
	\includegraphics[width = 0.5\textwidth]{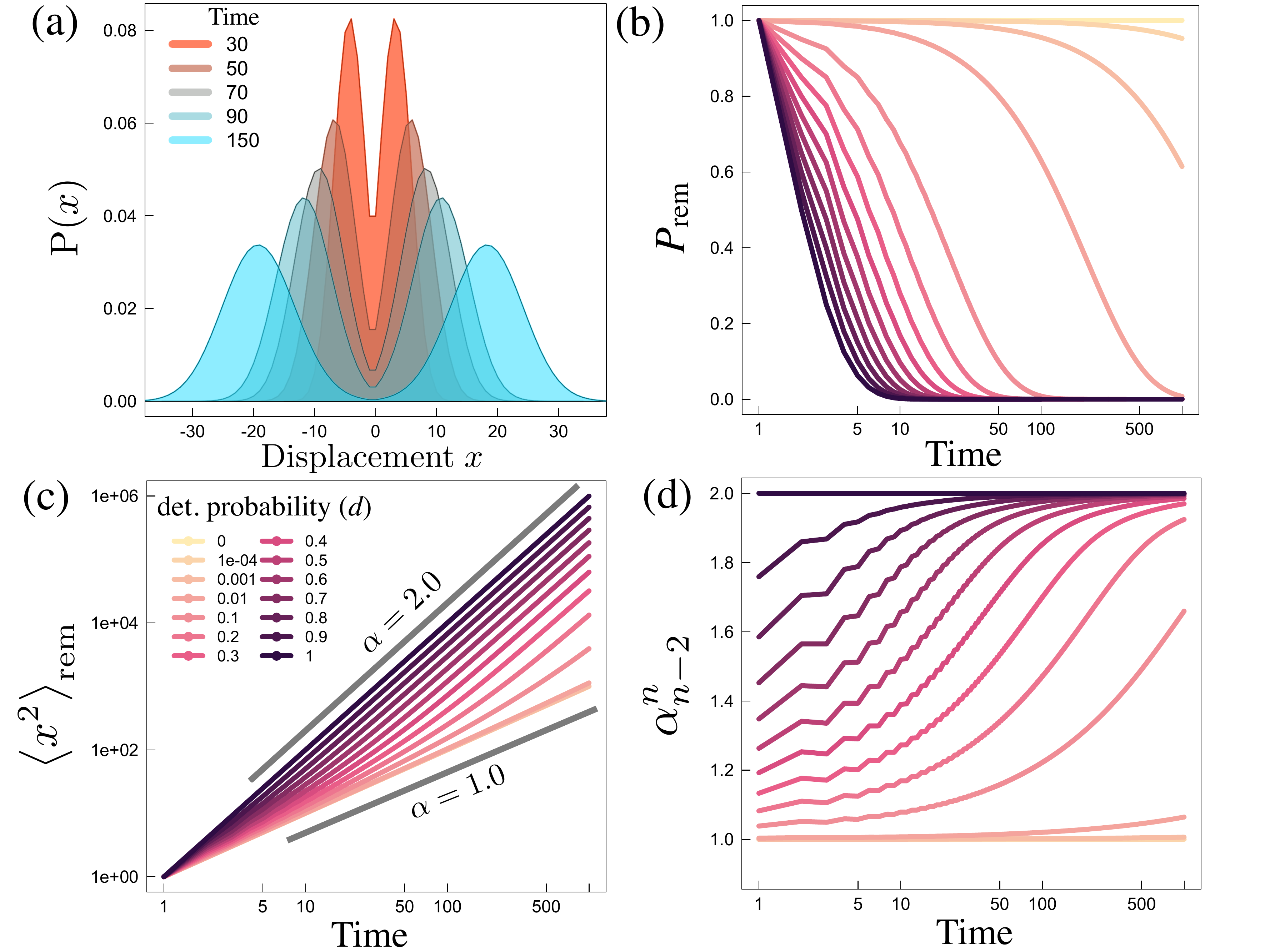}
	\caption{Ensemble-level analysis with constant detachment. (a) Evolution of the displacement distribution for constant-detachment probability $d = 0.4$. (b) Probability of remaining on the track as a function of time, for different detachment probabilities $d$.  
	(c) MSD$_{\rm EA}$~\eqref{rem} as a function of time.  (d) Diffusion exponent $\alpha^{n}_{n-b}$~\eqref{alphaMAIN} as a function of time. For $d=0$ (no detachment), $\alpha^{n}_{n-2} = 1$ (conventional diffusion) at all times. As the detachment probability increases to $d = 1$, $\alpha^{n}_{n-2}$ increases to the ballistic limit of 2.}
	\label{peaks}
\end{figure}

We first determine the effects of a constant origin-directed detachment bias on the ensemble-level system. Figure~\ref{peaks}a displays the evolution of the displacement distribution for the ensemble-level analysis with constant-detachment probability $d = 0.4$. Two peaks form, symmetric about the origin, and move with increasing dispersion in the $\pm \hat{x}$ directions, respectively. 
This is expected, as steps towards the origin have a probability of detachment, whereas steps away from the origin do not. 
Figure~\ref{peaks}b shows that as the probability $d$ of detachment increases, the probability of remaining on the track decays more rapidly. The MSD$_{\rm EA}$~\eqref{rem} varies with $d$, ranging from a ballistic trend ($\alpha = 2$) for detachment probability $d = 1$ to conventional diffusion ($\alpha = 1$) for the no-detachment scenario ($d = 0$) (Fig.~\ref{peaks}c). For intermediate $d$ ($0 < d < 1$), the MSD displays increasing superdiffusive trends with increasing $d$.

To more quantitatively extract the time dependence of the anomalous diffusion exponent, we calculate $\alpha^{n}_{n-2}$ using~\eqref{alphaMAIN}. For all intermediate values of $d$, Fig.~\ref{peaks}d shows that $\alpha$ increases with time monotonically towards the ballistic threshold; the transition to the ballistic threshold occurs sooner as $d \rightarrow 1$.

\begin{figure}[h]
	\includegraphics[width = 0.5 \textwidth]{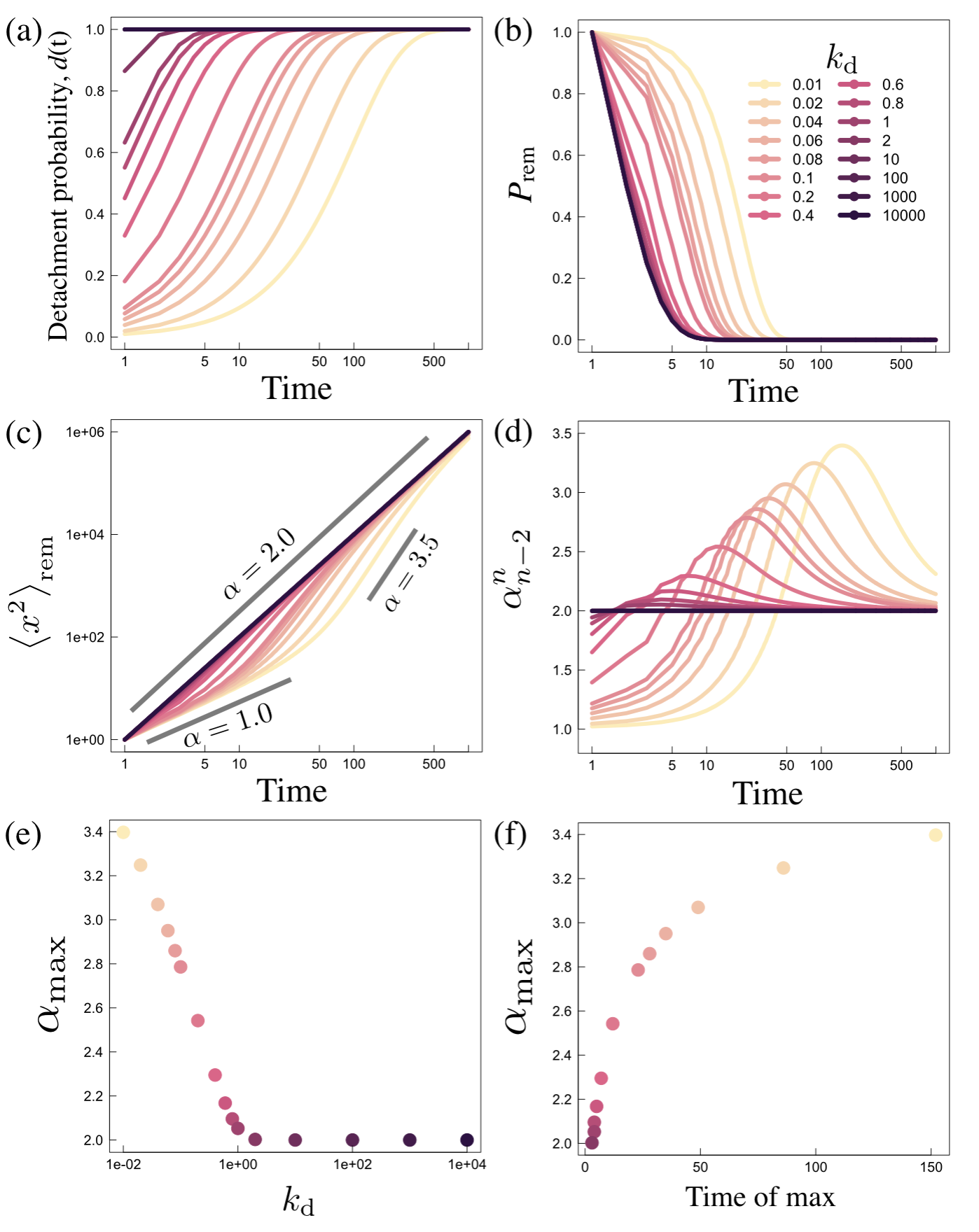}
	\caption{
	Ensemble-level analysis with exponential detachment. (a) Time-dependent detachment probability~\eqref{det} as a function of time for various decay rates $k_{\rm d}$. 
	(b) Probability of remaining as a function of time for various $k_{\rm d}$.
	(c) MSD$_{\rm EA}$~\eqref{rem} as a function of time, for various $k_{\rm d}$.
	(d) Transient behavior of $\alpha^{n}_{n-2}$~\eqref{alphaMAIN} as a function of time, for various $k_{\rm d}$. 
	(e) Maximum diffusion exponent $\alpha_{\mathrm{max}}$ as a function of $k_{\rm d}$. 
	(f) Parametric plot of $\alpha_{\mathrm{max}}$ versus the time at which $\alpha_{\mathrm{max}}$ occurs.}
	\label{MSD_alpha}
\end{figure}

Figure~\ref{MSD_alpha} shows the ensemble-level analysis with an exponential time-dependent detachment probability given by \eqref{det}.  $d(t)$ is still bounded by $[0,1]$, but increases monotonically with time towards unity at a rate governed by $k_{\rm d}$ (Fig.~\ref{MSD_alpha}a). The slowest detachment process studied, corresponding to $k_{\rm d} = 10^{-2}$, leads to the slowest decay of $P_{\rm rem}$ from 1 to 0 (Fig.~\ref{MSD_alpha}b). For the fastest detachment process studied ($k_{\rm d} = 10^{4}$), the MSD$_{\rm EA}$~\eqref{rem}
appears ballistic at all times (Fig.~\ref{MSD_alpha}c). As $k_{\rm d}$ decreases, the MSD transitions from diffusive to ballistic, exhibiting a long superballistic transient at intermediate times (peaking at $\alpha \approx 3.5$ for $k_{\rm d} = 10^{-2}$).
We find that all $k_{\rm d} < 1$ lead to $\alpha^{n}_{n-2}$
rising past the ballistic threshold. This behavior is characterized in Fig.~\ref{MSD_alpha}d, in which is is clear that as $k_{\rm d}$ decreases, the maximum $\alpha^{n}_{n-2}$ ($\alpha_{\rm max}$) increases further into the superballistic regime.  
Figure~\ref{MSD_alpha}e shows that $\alpha_{\rm max}$ decreases from strongly superballistic as $k_{\rm d}$ increases, until for $k_{\rm d} > 1$, $\alpha^{n}_{n-b}$ reaches and remains near the ballistic limit.
Figure~\ref{MSD_alpha}f shows that $\alpha_{\rm max}$ occurs later as $k_{\rm d}$ increases. Therefore, as the origin-biased detachment process is slowed, the system takes longer to breach the ballistic threshold, but extends farther into the superballistic regime and stays superballistic for longer.

When an ensemble of particles moves at the ballistic threshold, their velocity is constant; exceeding the ballistic threshold requires an ensemble acceleration ($\left\langle a(t) \right\rangle > 0 $).  
In Appendix D we derive a general expression for the mean velocity $\left\langle v_{+}(t) \right\rangle$
of the positive half of the 
displacement distribution (see Fig.~\ref{peaks}a).
For the constant and exponential biased-detachment models,
\begin{align}
\label{constVel}
\left\langle v_{+} \right\rangle_{\rm const} &= \frac{d}{2 - d} \\
\label{expVel}
\left\langle v_{+} \right\rangle_{\rm exp} &= \frac{1 - e^{-k_{\rm d}t}}{1 + e^{-k_{\rm d}t}} \ .
\end{align}
\eqref{constVel} predicts constant velocity (no acceleration; ballistic motion) for constant-detachment probability ($d > 0$), while \eqref{expVel} predicts acceleration (superballistic motion) when the origin-biased detachment probability increases with time according to \eqref{det}.

\begin{figure}[h]
	\includegraphics[width = 0.45 \textwidth]{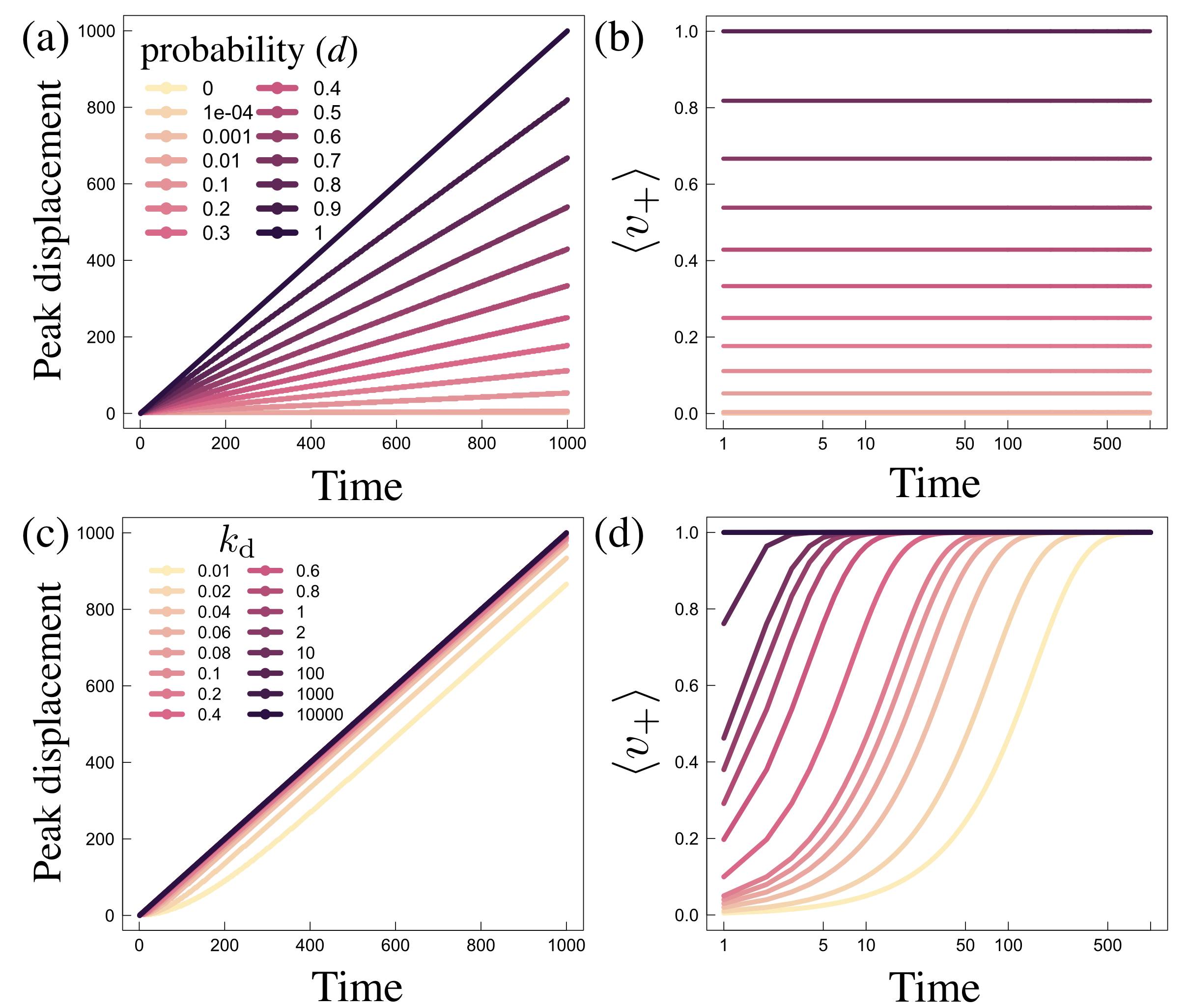}
	\caption{
	Ensemble-level analysis of positive peak displacement.
	(a) Positive peak displacement  
	and (b) its velocity~\eqref{constVel}, as a function of time for constant detachment. 
	(c) Positive peak displacement
	and (d) its velocity~\eqref{expVel} as a function of time for exponential detachment.
	}
	\label{mode}
\end{figure}

We compare these predictions with the dynamics of the 
ensemble-level displacement
distributions (Figs.~\ref{peaks},~\ref{MSD_alpha}).
For all constant-detachment probabilities, the positive peak displacement linearly increases with time (Fig.~\ref{mode}b).
The positive peak displacement for exponential detachment probability (Fig.~\ref{mode}c) increases nonlinearly, with velocity given by~\eqref{expVel}. 
In Appendix D, we validate~\eqref{expVel} by integrating it with respect to time to get the conditional mean displacement~\eqref{PeakD} and show that the analytical result agrees with the simulations (Fig.~\ref{AppenBFig}).

Despite our system moving with equidistant jump lengths, with equal probability in either direction, and at constant time intervals, the detachment protocol alone is sufficient to push the anomalous diffusion exponent far into the superballistic regime for hundreds of time steps. Therefore, despite no individual particle experiencing acceleration, the selective bias of the detachment protocol produces an ensemble acceleration sufficiently strong to push the diffusion exponent far into the superballistic regime. To the best of our knowledge, this is a new mechanism for anomalous diffusion.

\subsection{Trajectory-resolved analysis}

We have shown in an analytically tractable ensemble-level analysis that biased detachment is sufficient to tune the anomalous diffusion exponent far into the superballistic regime. 
Next, we explore the effects of the biased exponential detachment probability~\eqref{det} in a trajectory-resolved analysis.

\begin{figure}[h]
	\includegraphics[width = 0.5\textwidth]{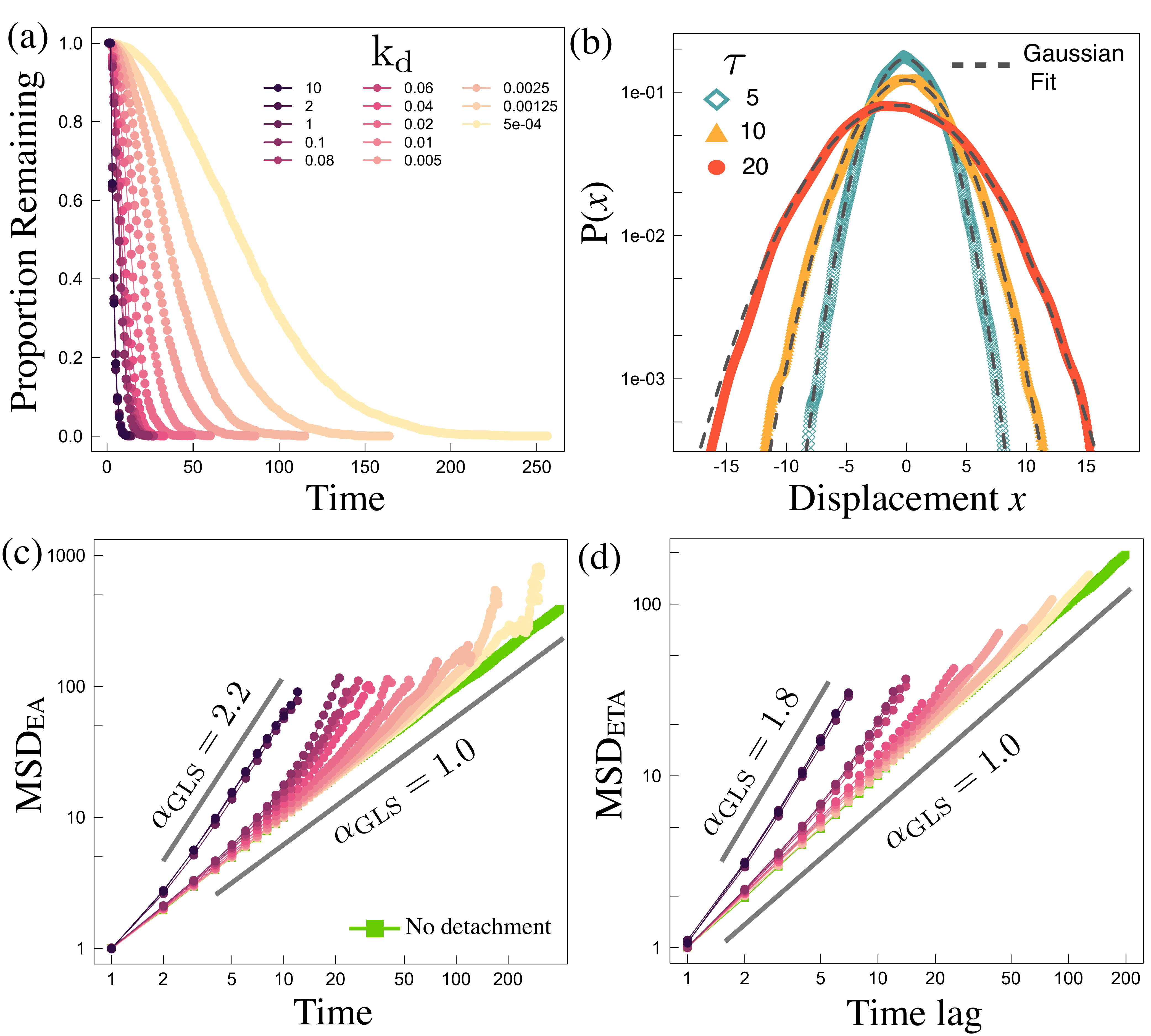}
	\caption{Trajectory-resolved analysis. (a) Proportion remaining as a function of time for various values of $k_{\rm d}$. (b) Example Gaussian fits to the displacement distribution for $k_{\rm d} = 0.005$ and time lags $\tau = 5,10,20$. (c) MSD$_{\rm EA}$~\eqref{EAMSD} as a function of time for each $k_{\rm d}$. (d) MSD$_{\rm ETA}$~\eqref{ETAMSD} as a function of time for each $k_{\rm d}$. 
    }
	\label{trajSpecAnalysis}
\end{figure}

For each $k_{\rm d}$, we simulate $5\times 10^{4}$ independent trajectories. Figure~\ref{trajSpecAnalysis}a displays the proportion of trajectories still continuing at different times:
as $k_{\rm d}$ increases, detachment happens more rapidly.

We use the Gaussian parameter $\gamma_2$~\eqref{kurt} to determine if the displacement distribution conforms to a Gaussian, which is thought to be a requirement for conventionally diffusive systems~\cite{Metzler2000}. For all $k_{\rm d}$, the displacement distribution is well approximated by a Gaussian distribution. As an example, Figure~\ref{trajSpecAnalysis}b shows the displacement distribution for time lags $\tau = 5$, 10, and 20 
for $k_{\rm d} = 0.005$. Figure~\ref{kurtosisAppen} in Appendix E shows that $\gamma_{2}$ is close to 0, the Gaussian limit, for these and other time lags.

Figures~\ref{trajSpecAnalysis}c and \ref{trajSpecAnalysis}d show the MSD$_{\rm EA}$~\eqref{EAMSD} and the MSD$_{\rm ETA}$~\eqref{ETAMSD} for all $k_{\rm d}$. The MSD$_{\rm EA}$ produces higher $\alpha_{\rm GLS}$ for every $k_{\rm d}$, with a maximum $\alpha_{\rm GLS} = 2.2$ for $k_{\rm d} = 10$, compared to $1.8$ from the MSD$_{\rm ETA}$ with the same $k_{\rm d}$. Whereas the MSD$_{\rm EA}$ breaks through the ballistic threshold, the MSD$_{\rm ETA}$ remains in the superdiffusive (sub-ballistic) regime. 
 
For the trajectory-resolved analysis the highest $k_{\rm d}$ we explore is 10, where $\alpha_{\rm GLS}$ peaks at 2.2 and 1.8 for the MSD$_{\rm EA}$ and MSD$_{\rm ETA}$, respectively (Fig.~\ref{trajSpecAnalysis}c,d). For the lowest $k_{\rm d}=5\times 10^{-4}$, $\alpha_{\rm GLS}$ barely increases above the no-detachment limit (Fig.~\ref{trajSpecAnalysis}c). 
This appears to be the opposite trend to that seen for the ensemble-level analysis (Fig.~\ref{MSD_alpha}e); however, the limited statistics of remaining trajectories means that the trajectory-resolved analysis covers a shorter time range than does the ensemble-level approach.

\begin{figure}[h]  
	\includegraphics[width = 0.45 \textwidth]{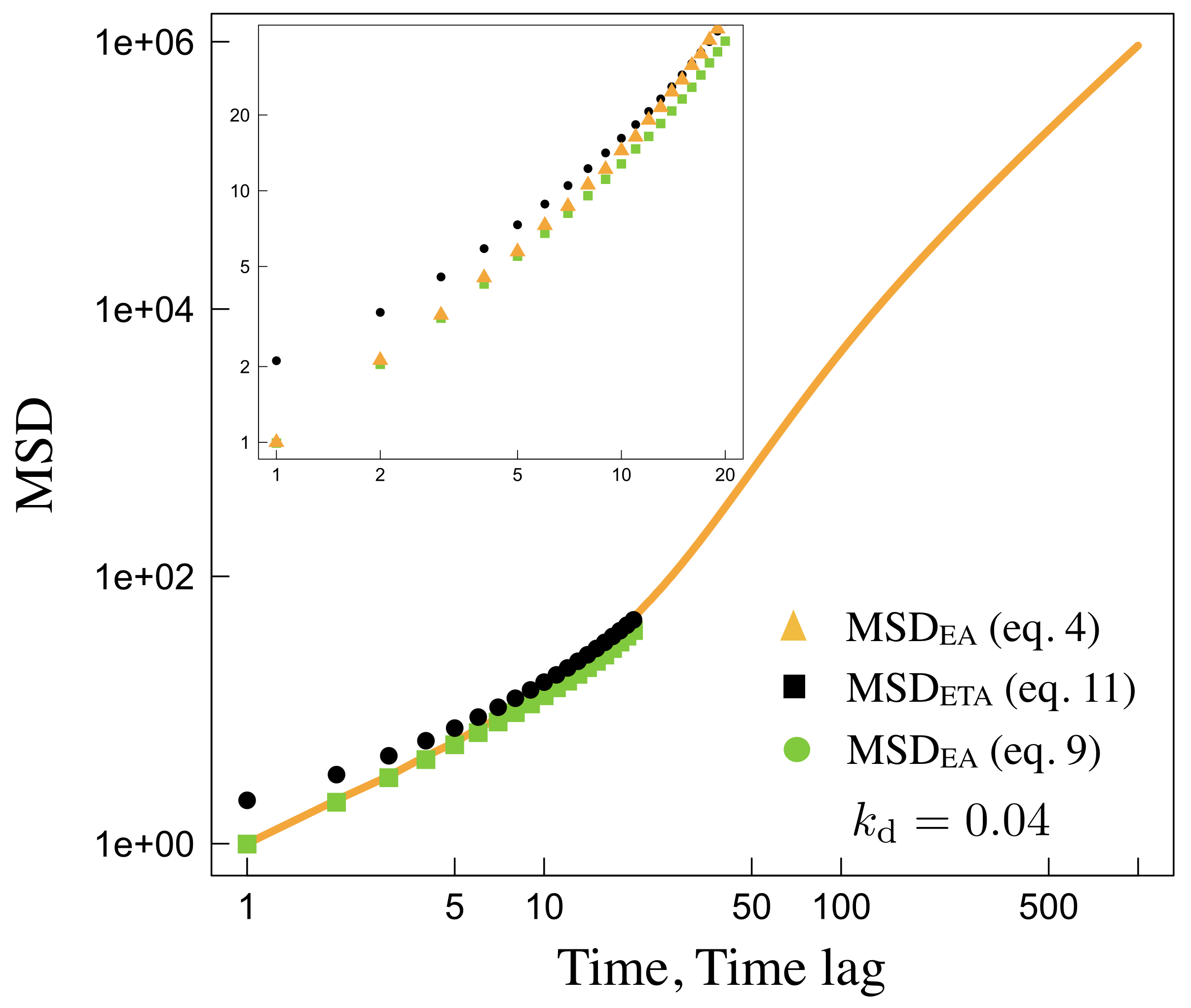}
	\caption{
	Trajectory-resolved MSD$_{\rm ETA}$~\eqref{ETAMSD} (black squares) and MSD$_{\rm EA}$~\eqref{EAMSD} (green circles) compared with ensemble-level MSD$_{\rm EA}$~\eqref{rem}	(orange triangles and line), for $k_{\rm d} = 0.04$ and 500,000 trajectories. Inset: all MSDs at short times or time lags. 
    }
	\label{MSDcomparison}
\end{figure}

To improve statistical significance, we simulate significantly more ($5\times10^5$) trajectories for $k_{\rm d} = 0.04$. 
Figure~\ref{MSDcomparison} shows trajectory-resolved MSD$_{\rm EA}$~\eqref{EAMSD} and MSD$_{\rm ETA}$~\eqref{ETAMSD}, alongside the ensemble-level MSD$_{\rm EA}$. This comparison makes clear that all MSDs follow the same trend. It also highlights the statistical challenge of achieving superballistic behavior by biased detachment; however, even a small detachment bias ($k_{\rm d} = 0.04$) produces significant superdiffusion. 

\subsection{Analysis of longest-duration trajectories}\label{subsection:longest}

Studies of molecular-motor transport commonly include in an MSD analysis only a subset of the recorded trajectories, often chosen because they remained associated to the track beyond a threshold duration~\cite{Salman2002, Howse2007,Manley2008} or reached a certain distance~\cite{Richard2019}. Here we show that with biased detachment, selecting a \textit{longest-lived} subset of the total ensemble biases the MSD analysis.

\begin{figure}[h] 
	\includegraphics[width = 0.5\textwidth]{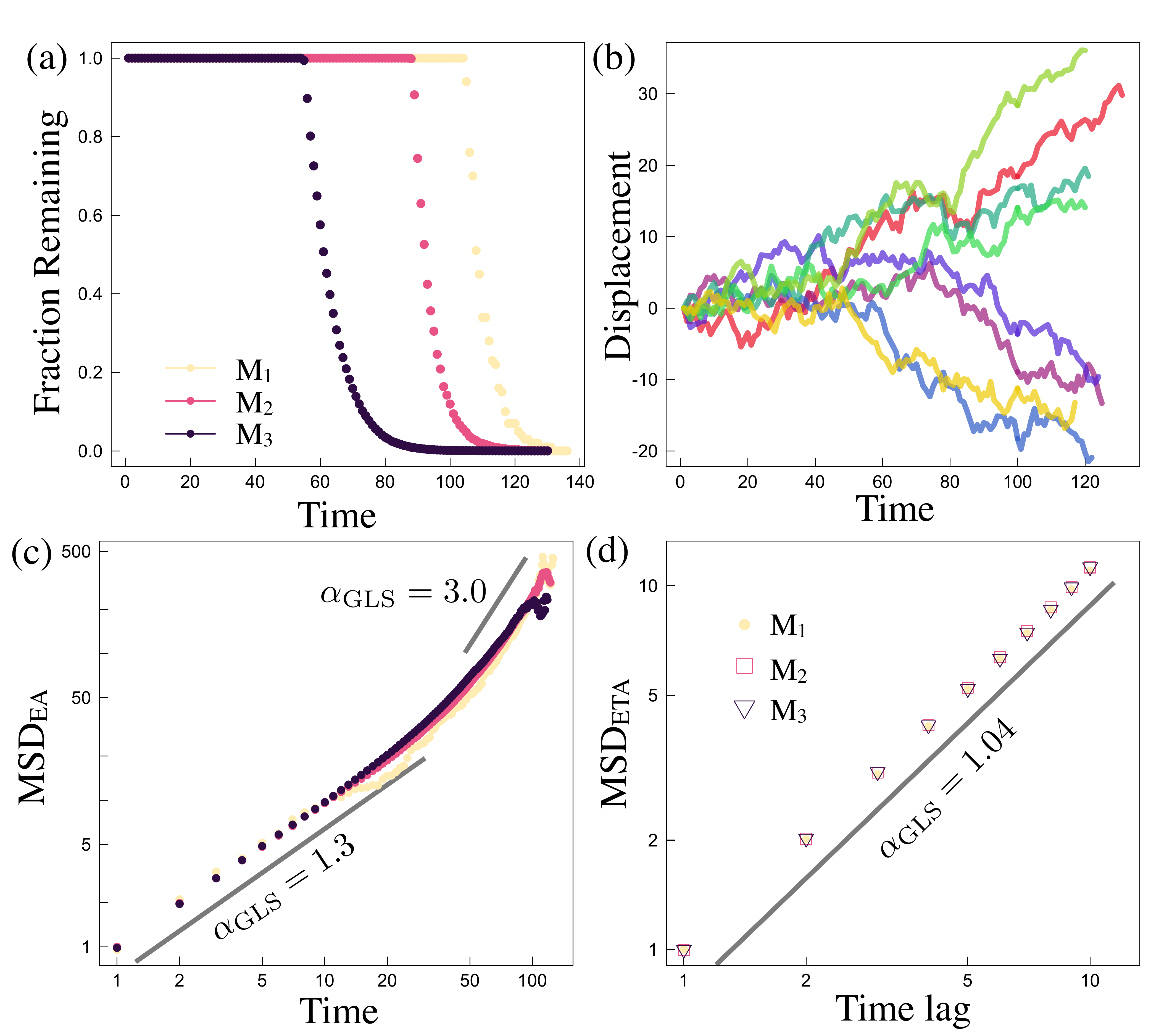}
	\caption{Longest-duration trajectory-resolved analysis. (a) Fraction remaining as a function of time for the $M$ longest-lived trajectories out of the total ensemble size of $2\times 10^{9}$. (b) Randomly selected trajectories from the longest-lived $M_{1}$ ensemble. 
	(c) MSD$_{\rm EA}$~\eqref{EAMSD} 
	and (d) MSD$_{\rm ETA}$~\eqref{ETAMSD} 
	as a function of time
	for all $M$ ensembles.
	All three ensembles give indistinguishable MSD$_{\rm ETA}$.
	}
	\label{longestlived1}
\end{figure}

We simulate $2\times 10^{9}$ independent trajectories with $k_{\rm d} = 0.05$, but only analyze subsets $M$ of these trajectories, selecting the $M_{1} = 10^2$, $M_{2} = 10^4$ and $M_{3} = 10^6$ longest-duration trajectories for each subset (corresponding to proportions $5\times 10^{-8}$, $5\times 10^{-6}$, and $5\times 10^{-4}$, respectively). 
Figure~\ref{longestlived1}a shows the detachment characteristics of these subsets: each $M$ sub-ensemble experiences a lag time before detachment begins. As the subset size 
decreases, detachment begins later.
Figure~\ref{longestlived1}b shows randomly selected trajectories from the longest-duration $M_{1}$ ensemble.
Despite their clear stochasticity, they exhibit net motion away from the origin (as expected for these trajectories that avoided early origin-biased detachment). 

Figure~\ref{longestlived1}c displays MSD$_{\rm EA}$ as a function of time for all three $M$ ensembles. At early times (where no trajectories detach) the diffusion exponent $\alpha_{\rm GLS}=1.3$, exceeding conventional diffusion. After the onset of detachment, $\alpha_{\rm GLS}$ increases far into the superballistic regime and peaks at $3.0$, as measured by a generalized least squares fit of the $M_{1}$ ensemble over the time interval indicated by the black slanted line in Fig.~\ref{longestlived1}c. 
Figure~\ref{longestlived1}d shows MSD$_{\rm ETA}$ as a function of time, where---by contrast---$\alpha \approx 1$ for all sub-ensembles.

The longest-lived trajectories are those that have managed to escape the `detachment trap' by successively moving away from the origin. That is, although the motion is random, from a sufficiently large ensemble there will be walkers that appear superdiffusive. It is therefore expected that the longest-lived ensembles exhibit superdiffusive characteristics even before the onset of detachment, as shown in Fig.~\ref{longestlived1}c. There is a stark disparity between the MSD measures, with the MSD$_{\rm EA}$ strongly superdiffusive ($\alpha = 1.3$, 
Fig.~\ref{longestlived1}c) and the MSD$_{\rm ETA}$ only slightly superdiffusive ($\alpha = 1.04 \pm 0.01$, Fig.~\ref{longestlived1}d). These results suggest that selecting a subset of trajectories based on processivity has the potential to strongly bias inferences about the system dynamics.

\subsection{Application to burnt-bridges ratchets}

Some synthetic biomolecular systems, such as molecular spiders~\cite{Pei2006, Lund2010},
burnt-bridges ratchets~\cite{Kovacic2015} and DNA nanomotors~\cite{Yehl2015, Blanchard2019, Salaita2020}, remodel their tracks as they move and have been engineered to achieve directional motion at the molecular level.
This remodeling turns a substrate site into a product site, and where there is a greater affinity to bind to substrate, motion is biased away from the product wake. 
Therefore, such systems have an increased probability of detachment from their tracks if they move backwards (into their product wake)~\cite{Samii2010, Samii2011, Olah2013, Korosec2018}. 

\begin{figure}[h]
	\includegraphics[width = 0.45 \textwidth]{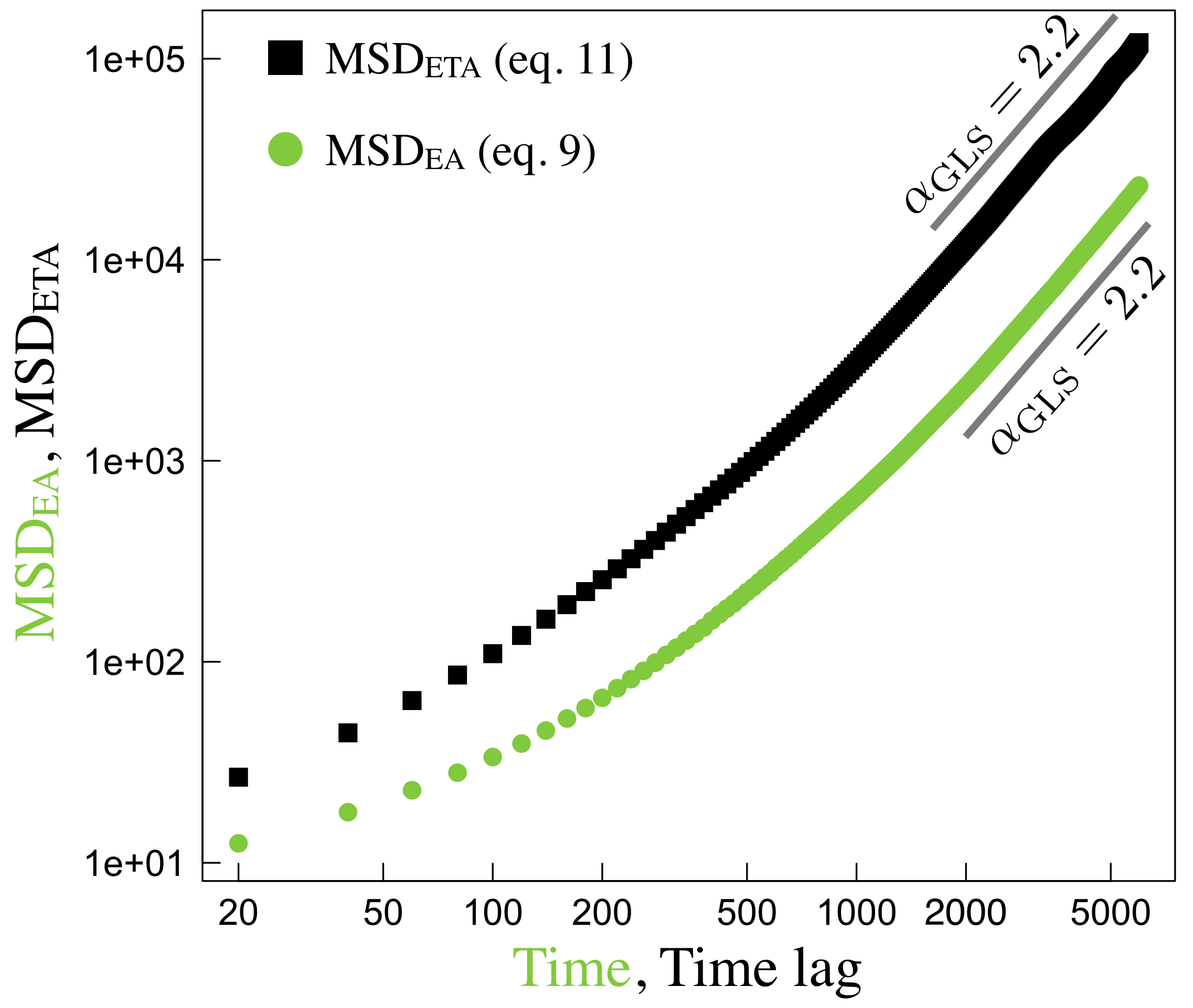}
	\caption{MSD$_{\rm EA}$ as a function of time (green) and MSD$_{\rm ETA}$ as a function of time lag (black) for burnt-bridges ratchets with finite processivity~\cite{Korosec2018}.}
	\label{BBRMSDs}
\end{figure}

To demonstrate that the anomalous $\alpha$ shown above is observed in more realistic systems with finite processivity, we examine simulated trajectories of a burnt-bridges ratchet (BBR), reported previously~\cite{Korosec2018}. Here, we examine an ensemble of 10,000 independent BBRs each moving on a quasi-1D track that is 4 lattice sites wide. Each BBR has 3 catalytic legs with a span of 8 lattice sites, which can each interact with substrate sites but not product sites.
Figure~\ref{BBRMSDs} shows that the MSD$_{\rm EA}$~\eqref{EAMSD} and MSD$_{\rm ETA}$~\eqref{ETAMSD} are quantitatively different. This is not surprising, because initial symmetry breaking leads to distinct short-time and long-time dynamics. 
In the long-time limit, both the MSD$_{\rm EA}$~\eqref{EAMSD} and MSD$_{\rm ETA}$~\eqref{ETAMSD} produce superballistic $\alpha_{\rm GLS}$ (2.19 and 2.17, respectively). As no acceleration is imposed on the system, the breach above the ballistic threshold likely arises, as above, from the origin-directed detachment bias inherent to the BBR dynamics. We note that this apparent superballistic behavior is for a specific example of BBRs described in \cite{Korosec2018} and may not be a general result for all BBR systems.

\section{Implications}
\subsection{Biological molecular motors}

MSD analysis is often utilized in investigations of molecular motor dynamics.
It has been applied to motors such as kinesin~\cite{Inoue2001}, dynein~\cite{Ross2006} and myosin~\cite{Levi2009}.
Of relevance to this work, all of these motors exhibit finite processivity: no molecular motor remains bound to its track forever. This finite processivity was the inspiration for the longest-lived analysis of Sec.~\ref{subsection:longest}: could duration-based selection of trajectories alter the conclusions drawn from MSD analysis? For example, when calculating the MSD$_{\rm TA}$~\eqref{TAMSD} from experimental data it is common to include only those motors that remained bound to the track for a certain time~\cite{Salman2002, Howse2007,Manley2008}, thereby ignoring those that detached. 
The selection of trajectories for MSD analysis is not always described,
which makes it difficult for the interested reader to assess any bias that may manifest in the MSD. (We expect ensemble selection is not typically described in detail because the resulting potential bias in the MSD has not been raised before.)
To our knowledge there are no experimental
studies that examine how the MSD changes as a function of the duration of the trajectories chosen. Such sub-sampling would demonstrate the robustness of the conclusions drawn from the MSD, and could potentially provide mechanistic insight about detachment kinetics. For example, sub-sampling may indicate a directional detachment bias, as suggested by the work presented in this manuscript.

\subsection{Random walks with finite processivity}

Much attention has focused on understanding systems displaying anomalous diffusion. Anomalous diffusion is often correlated with non-Gaussian displacement distributions~\cite{JosephPhillies2015,Wang2012}. By contrast, here we demonstrate that a system with a Gaussian displacement distribution (Fig.~\ref{trajSpecAnalysis}b) can exhibit strongly anomalous diffusion.
When simulating individual trajectories, our displacements are drawn from a Gaussian distribution.
Anomalous diffusion is achieved
by the consequences of the step direction: the walker faces an enhanced probability of detachment only if its step is towards the origin. Thus, the trajectories that tend towards the origin are filtered out as time progresses, 
thereby
splitting 
the
central mode into two symmetric modes about the origin (Fig.~\ref{MSD_alpha}). However, the displacement distribution for the entire ensemble is Gaussian (Fig.~\ref{trajSpecAnalysis}b) because the stepping dynamics are inherently Gaussian, as in regular diffusion. The distinction giving rise to superdiffusion (and even, at times, superballistic characteristics) is the preferential truncation of trajectories that step towards the origin.

\section{Conclusions}

The MSD is commonly used to assess anomalous diffusion in microscopic systems. 
Here we assessed the effect of a biased detachment probability on the MSD of one-dimensional random walkers, employing both constant and exponential detachment probabilities. 
We found that the detachment rate controls the apparent dynamics of the system: more gradual detachment (smaller $k_{\rm d}$) delays the onset of superballistic behavior, but results in a larger and longer-lasting excursion into the superballistic regime (Fig.~\ref{MSD_alpha}d).
All of these biased-detachment systems eventually relax to the ballistic limit ($\alpha = 2$). Fig.~\ref{MSDcomparison} shows a comparison between the trajectory and ensemble approaches where the superballistic behavior is most clearly seen in the (much longer-time) ensemble-level model. The limited statistics arising from detachment of trajectory-resolved systems means they cannot easily provide measures of long-time superballistic behaviors; however, the statistics are sufficient to demonstrate a strong superdiffusive bias.

The detachment bias confounds our ability to infer dynamical properties: the MSD suggests highly anomalous behavior, while the underlying dynamics we have imposed are Brownian. Therefore, for systems with trajectories of varying duration, a superdiffusive MSD should not on its own be taken to demonstrate directionality of a walker; such insight would require deconvolving the effects of processivity.

Researchers have cautioned about over-interpretation of the MSD for continuous-time random walks with power-law waiting-time distributions, where improper averaging can lead to false conclusions about transport properties~\cite{Lubelski2008}. We further caution the use of the MSD for systems with finite processivity, where detachment bias may lead to an overestimate of the anomalous diffusion exponent.

\section{Acknowledgements}
  
This work was funded by the Natural Sciences and Engineering Research Council of Canada (NSERC) through Discovery Grants to NRF and DAS and a Postgraduate Scholarship--Doctoral to CSK, and by a Tier-II Canada Research Chair to DAS. Computational resources were provided by Compute Canada.

\section*{Appendix A: Analytical expression for diffusion exponent $\alpha$}

The ensemble-averaged mean-squared displacement is defined as
\begin{equation} \label{n}
\left\langle [r(n)-r(0)]^{2} \right\rangle = \frac{1}{N}\sum_{i = 1}^{N}[r_{i}(n) - r_{i}(0)]^{2} = D_{\rm g}n^{\alpha}.
\end{equation}
Here, $n$ is an arbitrary point in time, $N$ is the size of the ensemble, $D_{\rm g}$ is the general diffusion coefficient and $\alpha$ is the anomalous diffusion exponent. We can also consider a point $n-b$ somewhat earlier in time:
\begin{equation} \label{n-b}
\left\langle [r(n-b) - r(0)]^{2} \right\rangle = D_{\rm g}(n-b)^{\alpha}.
\end{equation}
The ratio of 
\eqref{n} and \eqref{n-b} gives
\begin{equation} 
\frac{\left\langle [r(n) - r(0)]^{2} \right\rangle}{\left\langle [r(n-b)-r(0)]^{2} \right\rangle} = \frac{n^{\alpha}}{(n-b)^{\alpha}},
\end{equation}

\begin{align}
\begin{split}\label{eq:1}
\log_{n} \frac{\left\langle [r(n) - r(0)]^{2} \right\rangle}{\left\langle [r(n-b) - r(0)]^{2} \right\rangle} ={}& \log_{n} \frac{n^{\alpha}}{(n-b)^{\alpha}}
\end{split}\\
\begin{split}\label{eq:2}
={}& \alpha\log_{n} n \\
& - \log_{n}[(n-b)^{\alpha}].
\end{split}
\end{align}

Using the logarithmic identity
\begin{equation}\label{identity}
\log_{b} x = \frac{\log_{d} x}{\log_{d} b}
\end{equation}
we then solve for $\alpha$, 
\begin{equation}
\log_{n} \frac{\left\langle [r(n)-r(0)]^{2} \right\rangle}{\left\langle [r(n-b) - r(0)]^{2} \right\rangle} = \alpha \left( 1 - \frac{1}{\log_{n-b} n}\right) ,
\end{equation}
\begin{equation}\label{alpha}
\alpha^{n}_{n-b} = \left( 1 - \frac{1}{\log_{n-b} n}\right)^{-1} \log_{n} \frac{\left\langle [r(n)-r(0)]^{2} \right\rangle}{\left\langle [r(n-b)- r(0)]^{2} \right\rangle} \ . 
\end{equation}
The $\alpha^{n}_{n-b}$ notation denotes that $\alpha$ is estimated by two points, at times $n$ and $n-b$. Equation~\eqref{alpha} is an expression for $\alpha$ derived from an arbitrary ensemble-averaged mean-squared displacement. In the work presented in this manuscript we restrict ourselves to one dimension and have a system for which $\left\langle x \right\rangle = 0$. Equation~\eqref{alpha} then reduces to
\begin{equation}\label{alphApp}
\alpha^{n}_{n-b} = \left( 1 - \frac{1}{\log_{n-b} n}\right)^{-1} \log_{n} \frac{\left\langle x^2(n) \right\rangle}{\left\langle x^2(n-b) \right\rangle} \ .
\end{equation}

In this work we apply \eqref{alphApp} with $b = 2$ to analytical ensemble-level results, to estimate the anomalous diffusion exponent over a duration of 2 timesteps.

\section*{Appendix B: Comparing the dynamic functional to newly derived $\alpha_{n-b}^{n}$}

\begin{figure}[h!]
	\includegraphics[width = 0.45 \textwidth]{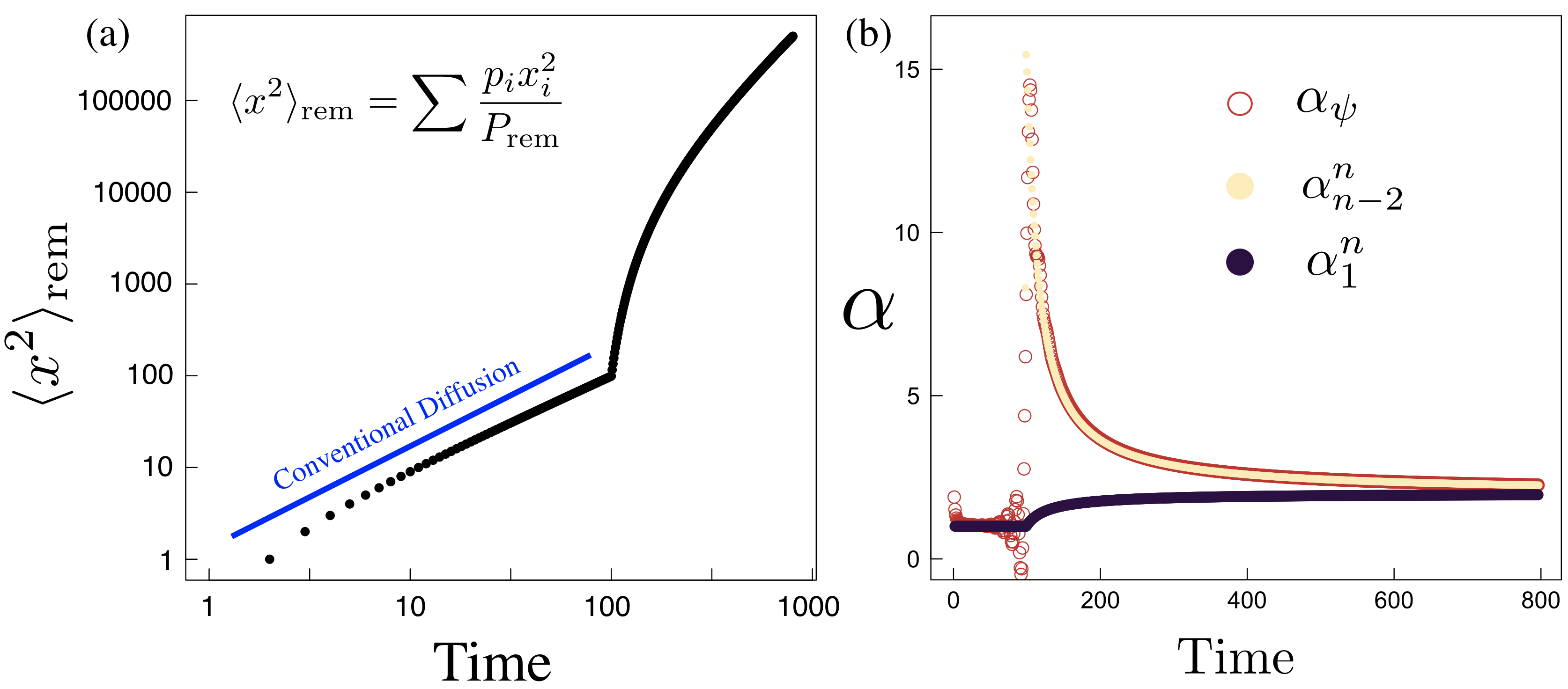}
	\caption{a) An example of a log-log MSD$_{\rm EA}$-time trend from the analytical model where detachment is turned on at $t$=100. b) A comparison of $\alpha$ computed numerically by \eqref{dynamicF} and analytically by \eqref{alphaMAIN}.
	}
	\label{comparison}
\end{figure}

To demonstrate the utility of $\alpha_{n-b}^{n}$ \eqref{alphaMAIN}, we determine
the anomalous diffusion coefficient 
for an example in which biased detachment is suddenly turned on after 100 timesteps. Figure~\ref{comparison}a shows conventional diffusion for the first 100 timesteps, after which there is a sudden increase in the MSD$_{\rm EA}$. Figure~\ref{comparison}b compares $\alpha$ determined using
\eqref{alphaMAIN} and \eqref{dynamicF}. The derivative used to calculate $\alpha_{\psi}$ via \eqref{dynamicF} is a smoothing spline derivative with cross-validation~\cite{Wahba1983}, while the expression \eqref{alphaMAIN} for $\alpha_{n-b}^{n}$ is exact.

\section*{Appendix C: Analytical calculations for the discrete random walk with detachment}

Table~\ref{table2} provides sample analytical calculations from Table~\ref{table1} 
of the ensemble-level MSD$_{\rm EA}$~\eqref{rem} and the analytical anomalous diffusion exponent $\alpha^{n}_{n-2}$ derived in Appendix A, up to 4 timesteps.
Here, $d$ is the probability of detachment~\eqref{det}, and $r \equiv 1 - d$ represents the probability of remaining attached to the lattice for an individual step. $P_{\rm rem}$ is a normalization factor and represents the total probability of remaining attached to the lattice.

\begingroup
\renewcommand{\arraystretch}{2}
\begin{table}[h!]
	\centering
	\caption{$\left\langle x^{2} \right\rangle _{\rm rem}$ and $\alpha^{n}_{n-2}$ for the first 4 timesteps.}

	\begin{tabular}{|
			>{\columncolor[HTML]{EFEFEF}}c |c|c|c|c|}
		\hline
		Timestep & \cellcolor[HTML]{EFEFEF}$P_{\rm rem}$                   & \cellcolor[HTML]{EFEFEF}$\sum p_{i}x_{i}^{2}$ & \cellcolor[HTML]{EFEFEF}$\left\langle x^{2}\right\rangle_{\rm rem}$ & \cellcolor[HTML]{EFEFEF}$\alpha^{n}_{n-2}$     \\ \hline
		0        & 1                                                   & 0                                             & 0                                       & /                                    \\ \hline
		1        & 1                                                   & 1                                             & 1                                       & /                                    \\ \hline
		2        & $\frac{1}{2}(1 + r)$                                & 2                                             & $\frac{4}{1 + r}$                       & /                 \\ \hline
		3        & $\frac{1}{4}(1 + 3r)$                             & $\frac{3}{4}(3 + r)$                          & $\frac{9 + 3r}{1 + 3r}$                 & $\log_{3} \frac{9 + 3r}{1 + 3r}$ \\ \hline
		4        & $\frac{1}{8}$ + $\frac{1}{2}r$ + $\frac{3}{8}r^{2}$ &$ 2(1 + r)$                                      & $\frac{16}{1 + 3r}$                     & $2\log_{4}\frac{4(r+1)}{3r+1}$        \\ \hline
	\end{tabular}
	\label{table2}
\end{table} 
\endgroup

\eject

\section*{Appendix D: Derivation of $\left\langle v_{+} \right\rangle$ for the ensemble-level model}\label{AppenB}

Here we derive a general expression for 
the ensemble-level
$\left\langle v_{+} \right\rangle$,
the
mean velocity conditioned on the walker being at positive displacement.
We consider a discrete system in one dimension that can take steps to the left or right with step size $|\Delta x| = 1$ and $\Delta t = 1$. The probability of stepping left or right is equal. Let $d_{\rm l}(t)$ and $d_{\rm r}(t)$ represent the probability of detaching from the lattice for steps taken to the left and right, respectively. We then write the normalized probability at time $t$ of taking a step to the left and remaining attached as
\begin{equation}
P(\Delta x_{\rm left}) = \frac{1 - d_{\rm l}(t)}{2 - d_{\rm l}(t) - d_{\rm r}(t)}, 
\end{equation}
and the normalized probability of taking a step to the right and remaining attached as
\begin{equation}
P(\Delta x_{\rm right}) = \frac{1 - d_{\rm r}(t)}{2 - d_{\rm l}(t) - d_{\rm r}(t)}.
\end{equation}
Thus, the average displacement after one timestep is $\left\langle \delta x \right\rangle = [P(\Delta x_{\rm right}) - P(\Delta x_{\rm left})] \Delta x$.
Then
\begin{equation}
\left\langle v_{+} \right\rangle = \frac{\left\langle \delta x \right\rangle }{\Delta t} = \frac{d_{\rm l}(t) - d_{\rm r}(t)}{2 - d_{\rm l}(t) - d_{\rm r}(t)} \frac{\Delta x}{\Delta t}\ .
\end{equation}
The effects of detachment on the mean ensemble acceleration $\left\langle a_{+} \right\rangle = \frac{\mathrm d}{\mathrm{d}t}\left\langle v_{+} \right\rangle$ can then be determined, as can the conditional mean displacement: $\left\langle x_{+} \right\rangle = \int \mathrm{d}t \left\langle v_{+} \right\rangle$. 

In this work we consider exponential detachment for particles moving towards the origin: $d_{\rm r}(t) = 0$ and $d_{\rm l}(t) = d(t) = 1 - e^{-k_{\rm d}t}$. We then have 
\begin{equation}
{\left\langle v_{+} \right\rangle}_{\rm exp} =  \frac{1 - e^{k_{\rm d}t}}{1 + e^{-k_{\rm d}t}}.
\end{equation}
We also consider the case where detachment towards the origin is constant with time, $d_{\rm r}(t) = 0$ and $d_{\rm l}(t) = d$, giving
\begin{equation}
{\left\langle v_{+} \right\rangle}_{\rm const} =  \frac{d}{2 - d}.
\end{equation}

\begin{figure}[h!]
	\includegraphics[width = 0.45 \textwidth]{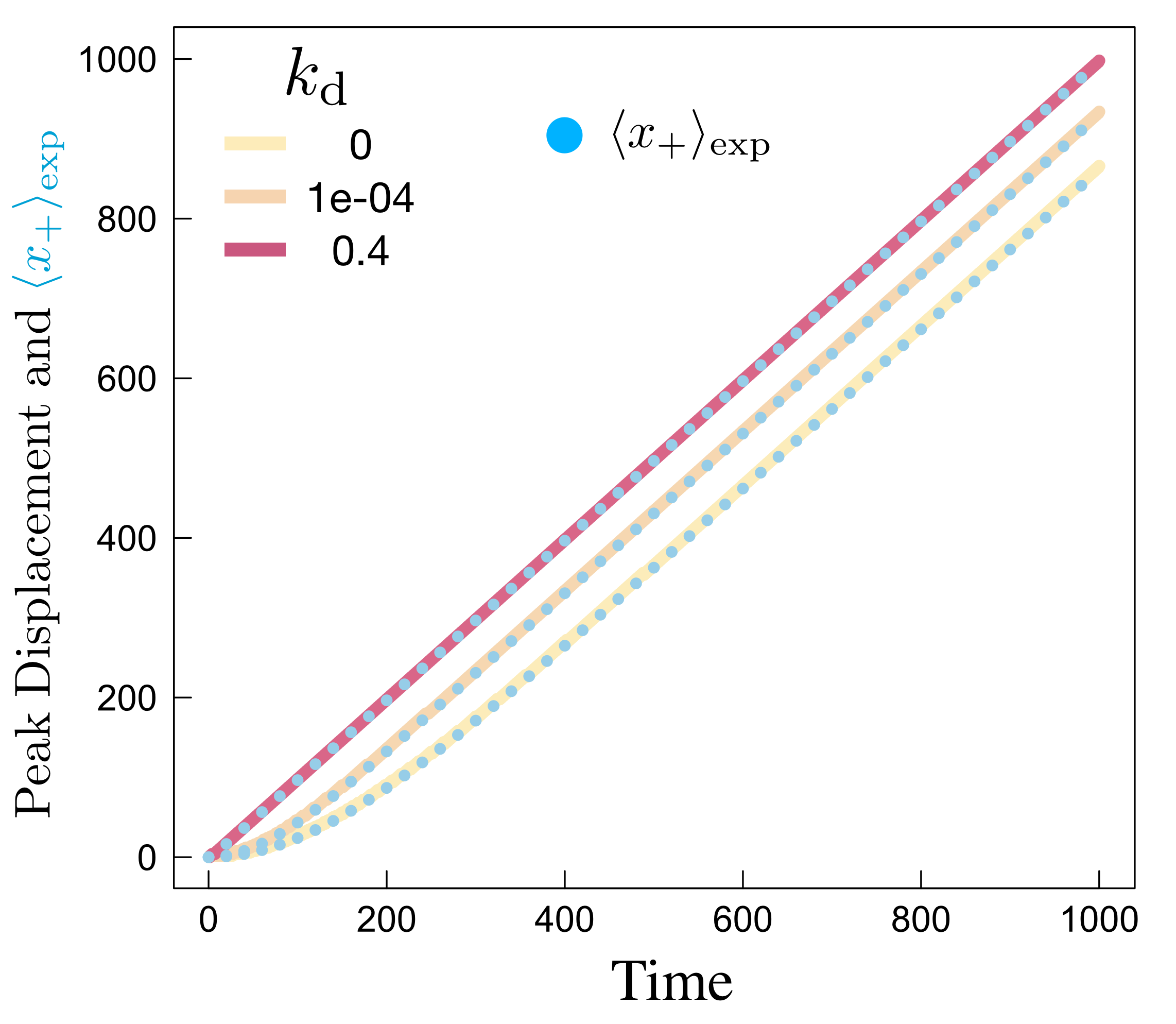}
	\caption{
	Conditional mean $\langle x_{+} \rangle_{\rm exp}$~\eqref{PeakD} (blue dots) from theory agrees with ensemble-level positive peak displacement (curves) from Fig.~\ref{mode}c.
    }
	\label{AppenBFig}
\end{figure}

To demonstrate agreement of this approach with our numerical data, we compare for the exponential case the positive peak displacement obtained from the ensemble-level simulations with the analytical conditional mean
\begin{align}\label{PeakD}
    {\left\langle x_{+} \right\rangle}_{\rm exp} ={}& \int \mathrm{d}t \frac{1 - e^{k_{\rm d}t}}{1 + e^{-k_{\rm d}t}} \\ ={}& \frac{2\log ({e^{k_{\rm d}t} + 1})}{k_{\rm d}} - t + C,
\end{align}
where $C$ is an integration constant. We set the integration constant to $C = -\frac{2\log 2}{k_{\rm d}}$ such that $\langle x_{+} \rangle_{\rm exp}(t = 0) = 0$. 
We find this analytical expression for $\langle x_{+} \rangle_{\rm exp}$ to agree with the ensemble-level results for exponential detachment (Fig.~\ref{AppenBFig}). 

\section*{Appendix E: Further Kurtosis Analysis of trajectory-resolved model}

Here we plot the excess kurtosis as a function of time lag for $k_{\rm d} = 0.005$ of the trajectory-resolved simulations. We find that at all time lags the excess kurtosis is close to the Gaussian limit of 0. 

\begin{figure}[h]
	\includegraphics[width = 0.45\textwidth]{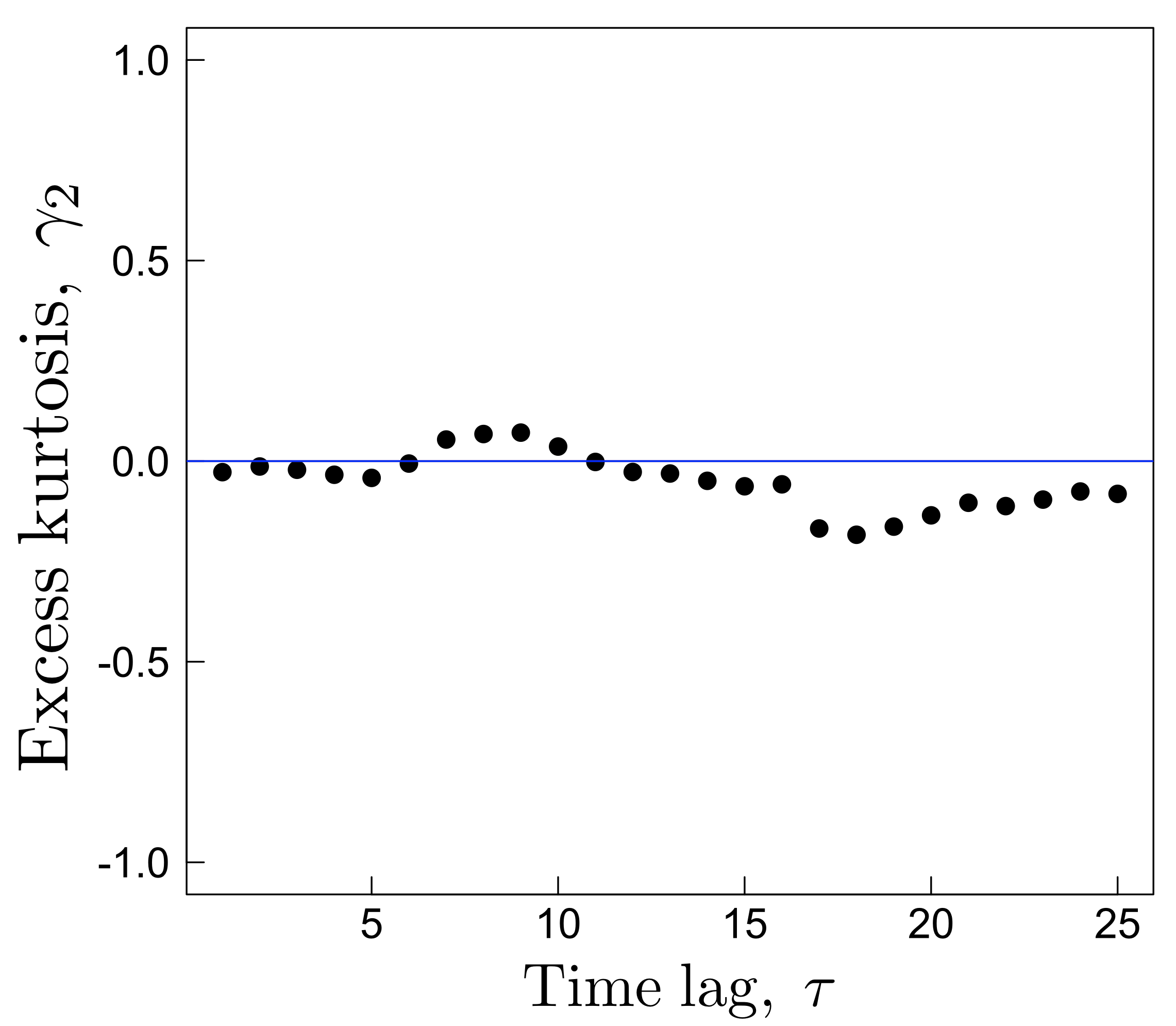}
	\caption{Excess kurtosis as a function of time lag for the trajectory-resolved model with $k_{\rm d} = 0.005$.}
	\label{kurtosisAppen}
\end{figure}

\eject

%

\end{document}